\documentclass{Interspeech2024}
\usepackage{booktabs} 
\usepackage{tabularx} 
\usepackage{graphicx} 
\usepackage{booktabs}
\usepackage{longtable}
\usepackage{lipsum}
\usepackage{stfloats} 
\usepackage{authblk}
\usepackage{multicol}
\usepackage{setspace}
\usepackage{multirow}

\interspeechcameraready

\title{What Does it Take to Generalize SER Model Across Datasets? A Comprehensive Benchmark}

\name[affiliation={1}]{Adham}{Ibrahim}
\name[affiliation={1,2}]{Shady}{Shehata}
\name[affiliation={1,3}]{Ajinkya}{Kulkarni}
\name[affiliation={1}]{Mukhtar}{Mohamed}
\name[affiliation={1,2,4}]{Muhammad}{Abdul-Mageed}
\address{
  $^1$Mohamed bin Zayed University of Artificial Intelligence, UAE; 
  $^2$Invertible AI, UAE;
  $^3$IDIAP, Switzerland;
  $^4$University of British Columbia, Canada
  }

\email{ \{adham.ibrahim, shady.shehata\}@mbzuai.ac.ae}


\begin{document}

\maketitle

\begin{abstract}

\
Speech emotion recognition (SER) is essential for enhancing human-computer interaction in speech-based applications. Despite improvements in specific emotional datasets, there is still a research gap in SER's capability to generalize across real-world situations. In this paper, we investigate approaches to generalize the SER system across different emotion datasets. In particular, incorporate 11 emotional speech datasets and illustrate a comprehensive benchmark on the SER task. We also address the challenge of imbalanced data distribution using oversampling methods when combining SER datasets for training. Furthermore, we explore various evaluation protocols for adeptness in the generalization of SER. Building on this, we explore the potential of Whisper for SER, emphasizing the importance of thorough evaluation. Our approach is designed to advance SER technology by integrating speaker-independent methods. 

\end{abstract}

\section{Introduction}

Emotions are intrinsic characteristics of human communications expressed through speech modality and plays crucial role in human-machine interfaces \cite{Kulkarni2022ExpressivityTI}. 
Over the past decades, research has been focused on various feature extraction methods along with machine learning. This includes traditional approaches like Mel-frequency cepstral coefficients (MFCCs)\cite{MFCC_SER}, linear prediction cepstral coefficients (LPCCs)\cite{LPCCs_SER}, and prosodic features\cite{prosodic_features}. Deep learning techniques such as deep neural network DNN\cite{DNN_SER}, recurrent neural network (RNN)\cite{RNN_SER}, and convolutional neural networks (CNN)\cite{CNN_SER1} have also gained prominence. The last three years have seen a considerable exploration of HuBERT\cite{hsu2021hubert} and Wav2Vec2\cite{IEMOCAP_NEW1} models in SER \cite{dawn_Transformer_SER, IEMOCAP_NEW1}, demonstrating their potential for robust emotion recognition.

IEMOCAP \cite{busso2008iemocap} is considered the benchmark dataset for discrete speech emotion recognition. An extensive evaluation of SER systems has been conducted using IEMOCAP, employing various machine learning methods such as support vector machines (SVM), LSTM, CNN, and ensemble learning \cite{IEMOCAP_BENCHMARK_OLD1, IEMOCAP_BENCHMARK_OLD2}. However, the use of Self-supervised learning (SSL) speech models (e.g., Hubert \cite{hsu2021hubert} and Wav2Vec2 \cite{IEMOCAP_NEW1}) has emerged as the state-of-the-art approach in SER, producing the best performance on the IEMOCAP dataset \cite{IEMOCAP_NEW1}. Another very important acted dataset is (RAVDESS) \cite{livingstone2018ryerson_RAVDESS}. RAVDESS was also involved in benchmarking many different SER systems \cite{ravdess_banchmark1, combination_survey}. A dataset closely similar to RAVDESS is CREMA-D \cite{cao2014crema}, which comprises 12 distinct spoken sentences performed by 91 different actors, resulting in a total of 7442 unique audio files.

Among the more compact SER databases available, the Surrey Audio-Visual Expressed Emotion (SAVEE) \cite{jackson2014surrey_savee} and the Toronto Emotional Speech Set (TESS) \cite{pichora2010toronto_TESS} stand out. Both datasets have only been benchmarked using traditional machine learning techniques (SVM, LSTM, CNN) \cite{Savee_TESS_benchmark1, Savee_TESS_benchmark2, Savee_TESS_benchmark4}. To the best of our knowledge, these datasets have yet to be tested with newer transformer-based models. MELD dataset has a unique conversational setup of 13,000 utterances from 1,433 dialogues from the TV series Friends. This compilation aims to address the complexities of emotion recognition within conversational contexts, a task known for its challenges.
Other SER datasets include ASVP-ESD\cite{tientcheu_touko_landry_dejoli_2021_4782712_ASVP-ESD}, EmoV-DB\cite{adigwe2018emotional_Emov_DB},  EmoFilm\cite{emilia_parada_cabaleiro_2018_1326428_EmoFilm},  JL-Corpus\cite{james2018open}, and ESD\cite{zhou2022emotional}.\\
Despite the availability of various datasets for speech emotion recognition, there's a consensus that SER still trails behind other tasks  \cite{Liu2020TemporalAC, 8462685, bao19_interspeech}. This shortfall is attributed to the absence of a large, universal dataset capable of bridging the significant disparities among existing datasets and facilitating adaptation to real-world scenarios. A recent study \cite{combination_survey} investigated the generalization capabilities and the possibility of merging different SER datasets by applying a set of cross-validation experiments, considering both single datasets and combinations of them, on RAVDESS, TESS, CREMA-D, and IEMOCAP. Their results confirm that SER models do not generalize well across datasets (training and testing on different datasets) and that merging datasets can mitigate this problem, as SER performance improves with access to a larger and more varied collection of data points.\\
A key challenge in speech emotion recognition stems from the substantial variance among available datasets, attributed to differences in setup (acted, natural, and elicited), recording quality, and subjective emotion perception by speakers and annotators. Consequently, most systems struggle to generalize across datasets, hindering performance in real-world applications \cite{combination_survey}. Moreover, emotional states influence spoken sentences, compounded by personal attributes like age and gender, posing obstacles for SER generalization and hindering the development of speech-independent systems, particularly in tasks like speech emotion recognition (SER). This challenge is underscored by human behavior, as demonstrated in an experiment by Schuller et al. \cite{Schuller} where the accuracy of 12 participants in recognizing expressed emotions decreased from 87.3\% with familiar speakers to 64.7\% when identifying emotions from speakers they had not previously encountered.\\
Due to variations in emotion classes, emotion labeling, amount of emotional data per dataset, SER more arduous task to generalize. In this paper, we present novel comprehensive benchmarking results across 11 SER datasets and various experimentations showcasing the vital aspects of generalization and performance improvement of SER.

\section{Methodology}

\begin{figure}[!t]
    \centering
    \includegraphics[width=0.4\textwidth]{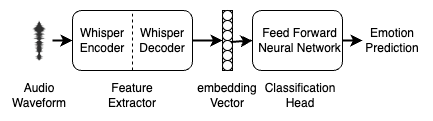}
    \caption{Whisper based speech emotion recognition system}
    \label{fig:arch}
\end{figure}

In this section, we highlight our primary contributions: Our SER system based on Whisper, the combination of different datasets, and evaluation strategy.

\subsection{Model Architecture}




We illustrate the SER model architecture, which we use throughout this work for various experiments across various SER datasets. This architecture aligns with the current state-of-the-art (SoTA) and is representative of the current best practices in speech emotion recognition~\cite{dawn_Transformer_SER}.\\
Our SER first uses Whisper-based feature extraction~\cite{Whisper} to map speech into latent representation through a transformer-based encoder-decoder network as shown in Fig 1. We use the base version of Whisper utilizing both the encoder and decoder to extract fixed-size embedding and fine-tune the whole model with the rest of the network. Thereafter, this latent representation as features is given to feed-forward of fully connected layers: starting from a size of 4096, it sequentially narrows down through layers of 2048, 1024, and 512, with each layer followed by a ReLU activation function to add non-linearity. This setup culminates in an output layer tailored to the number of target classes, making it suitable for classification purposes. To our knowledge, Whisper-based SER has not been explored for ASR tasks. We implement a basic five-layer feed-forward neural network, where the last layer projects into the different emotion labels. Cross-entropy serves as our loss function, and we maintain a constant learning rate of 0.00001. We train all variants of SER on NVIDIA A100-SXM GPUs equipped with 40GB of memory. 

\subsection{Datasets Preparation}
Emotional databases in speech emotion recognition are classified into acted, elicited, and natural categories, each with unique characteristics and challenges. Acted databases consist of recordings from trained actors simulating specific emotions, offering control but relying on the actors' skills\cite{Acted_datasets}. Elicited databases capture genuine emotional responses under designed scenarios requiring ethical consent, while natural databases compile authentic interactions from everyday life, presenting a wide authenticity spectrum but limited emotional range\cite{Elicit_datasets}.

Our study conducts a detailed analysis of individual and combined datasets to understand the impact of similar emotions on model accuracy and generalization on 11  datasets in the SER domain: \textbf{IEMOCAP}\cite{busso2008iemocap}: Approximately 12 hours from 10 speakers, with scripted and improvised dialogues across seven emotion classes (anger, happiness, sadness, excited, frustrated, and neutral). \textbf{MELD}\cite{poria2019meld}: 13,708 utterances over 12 hours from the Friends TV series, categorized into seven emotions (anger, disgust, sadness, joy, neutral, surprise, fear) with 10 main speakers and around 290 secondary speakers.  \textbf{ASVP-ESD}\cite{tientcheu_touko_landry_dejoli_2021_4782712_ASVP-ESD}: A multi-lingual corpus from movies, YouTube, and real interactions, focusing on English speech with seven emotions, totaling 6.5 hours from main and secondary speakers. \textbf{EmoV-DB}\cite{adigwe2018emotional_Emov_DB}: Covers five emotions with five speakers (four in English, one in French), totaling 9.5 hours. \textbf{TESS}\cite{pichora2010toronto_TESS}: Features two actresses, covering seven emotion classes over 2800 files in 1.6 hours. \textbf{EmoFilm}\cite{emilia_parada_cabaleiro_2018_1326428_EmoFilm}: Samples in English, Italian, and Spanish from 43 movies, focusing on five emotions over 20 minutes. \textbf{SAVEE}\cite{jackson2014surrey_savee}: British-English corpus with 480 utterances from four speakers, spanning seven emotions in 30 minutes. \textbf{RAVDESS}\cite{livingstone2018ryerson_RAVDESS}: Focuses on the speech part with 1440 utterances from 24 actors, covering eight emotions over 1.5 hours.
\textbf{CREMA-D}\cite{cao2014crema}: Audio-visual dataset with 7442 stimuli from 91 actors, spanning six emotions over 5.3 hours. \textbf{JL\_corpus}\cite{james2018open}: Contains 2400 sentences from four speakers, covering 10 emotions over 1.4 hours. \textbf{ESD}\cite{zhou2022emotional}: Offers 350 parallel utterances in five emotion categories from 10 native English speakers, with over 29 hours recorded.

\begin{table}[!t]
\setlength\tabcolsep{0.5pt} 
\caption{Distribution of emotion samples across 11 SER datasets}
\label{tab:combined_counts}
{\tiny
\begin{tabular}{lcccccccccccccc}
\toprule
\multirow{1}{*}{} & \multicolumn{11}{c}{\textbf{Dataset Name}} \\
\cmidrule(lr){2-12}
\textbf{Emotion} & \textbf{ASVP-ESD} & \textbf{CREMA-D} & \textbf{EmoF} & \textbf{ESD} & \textbf{IEMOCAP} & \textbf{JL} & \textbf{MELD} & \textbf{TESS} & \textbf{emov\_db} & \textbf{savee} & \textbf{ravdess} & \textbf{Total} \\
\midrule
\textbf{angry} & 585 & 1271 & 77 & 3500 & 1103 & 240 & 1607 & 400 & 1268 & 60 & 192 & 10303 \\
\textbf{disgust} & 612 & 1271 & - & - & - & - & 301 & 400 & 522 & 45 & 184 & 3335 \\
\textbf{anxious} & - & - & - & - & - & 240 & - & - & - & - & - & 240 \\
\textbf{apologetic} & - & - & - & - & - & 240 & - & - & - & - & - & 240 \\
\textbf{assertive} & - & - & - & - & - & 240 & - & - & - & - & - & 240 \\
\textbf{concerned} & - & - & - & - & - & 240 & - & - & - & - & - & 240 \\
\textbf{encouraging} & - & - & - & - & - & 240 & - & - & - & - & - & 240 \\
\textbf{excited} & - & - & - & - & 1041 & 240 & - & - & - & - & - & 1281 \\
\textbf{frustrated} & - & - & - & - & 1849 & - & - & - & - & - & - & 1849 \\
\textbf{fear} & 544 & 1271 & 72 & - & - & - & 358 & 400 & - & 60 & 192 & 2897 \\
\textbf{happy} & 652 & 1271 & 70 & 3500 & 595 & 240 & 2307 & 400 & - & 60 & 192 & 9287 \\
\textbf{neutral} & 714 & 1087 & - & 3500 & 1708 & 240 & 6432 & 400 & 1568 & 120 & 96 & 15865 \\
\textbf{pain} & 558 & - & - & - & - & - & - & - & - & - & - & 558 \\
\textbf{sad} & - & 1271 & 74 & 3500 & 1084 & 240 & 1002 & 400 & - & 60 & 192 & 7823 \\
\textbf{surprised} & 695 & - & - & 3500 & 107 & - & 1403 & 400 & - & 60 & 192 & 6357 \\
\textbf{contempt} & - & - & 50 & - & - & - & - & - & - & - & - & 50 \\
\textbf{amused} & - & - & - & - & - & - & - & - & 1317 & - & - & 1317 \\
\textbf{sleepy} & - & - & - & - & - & - & - & - & 1721 & - & - & 1721 \\
\textbf{calm} & - & - & - & - & - & - & - & - & - & - & 192 & 192 \\
\bottomrule
\end{tabular}
\label{table:dataset_counts}
}
\end{table}

In our investigation, we specifically extracted English speech and corresponding labels from mixed-language datasets like ESD\cite{zhou2022emotional}, ASVP-ESD\cite{tientcheu_touko_landry_dejoli_2021_4782712_ASVP-ESD}, and EmoFilm\cite{emilia_parada_cabaleiro_2018_1326428_EmoFilm}. It's essential to highlight that the other datasets employed in our study are exclusively in English, ensuring a consistent linguistic foundation for our analysis. Regarding the IEMOCAP dataset, our dataset preparation followed the guidelines outlined in a previous study\cite{10.1007/978-3-030-27535-8_43}, where emotions were categorized into anger, happiness, sadness, and neutral, with excitement categorized under happiness. We retained the original emotion classes for the remaining datasets. Additionally, we downsampled speech across all speech utterances to 16 kHz. Speaker identification for each audio clip in every dataset was accomplished using the information provided in the dataset metadata.

\subsection{Combining datasets for the SER task}

A major challenge in speech emotion recognition lies in the considerable variance among the available datasets. This variance stems from differences in their setup (acted, natural, and elicit), the quality of recordings, and the subjective perception of emotions by different speakers and annotators. Consequently, most speech emotion recognition systems do not generalize well across different datasets (training and testing on different datasets) \cite{combination_survey} and, by extension, struggle to perform well in real-world applications. 

In an effort to address this challenge, \cite{combination_survey} showed promising results by combining four of the most famous SER datasets (RAVDESS\cite{livingstone2018ryerson_RAVDESS}, TESS\cite{pichora2010toronto_TESS}, CREMA-D\cite{cao2014crema}, and IEMOCAP\cite{busso2008iemocap}). In this work, in addition to these four datasets, we added seven more (MELD\cite{poria2019meld}, ASVP-ESD\cite{tientcheu_touko_landry_dejoli_2021_4782712_ASVP-ESD}, EmoV-DB\cite{adigwe2018emotional_Emov_DB}, EmoFilm\cite{emilia_parada_cabaleiro_2018_1326428_EmoFilm}, SAVEE\cite{jackson2014surrey_savee}, JL-Corpus\cite{james2018open}, and ESD\cite{zhou2022emotional}) for a total of eleven SER datasets. To the best of our knowledge, combining 11 datasets presents the largest training dataset for the SER task. This study aims to significantly boost performance (against individual datasets) on the SER task, attributable to the considerably larger and more diverse array of training data points. 

\subsection{Evaluation protocol}

Throughout this work, we opted for the leave-one-speaker-out (LOSO) method to evaluate our models. We used accuracy as a way to measure the performance of SER systems. When dealing with datasets where all speakers contributed the same amount of speech, we randomly selected one speaker from each dataset as our test case. Conversely, for datasets where speakers had varying amounts of speech time, such as MELD, ASVP-ESD, and IEMOCAP, we excluded the speaker with the most speech duration. At the end of this process, we ensured that a speaker from every dataset was chosen, guaranteeing a wide-ranging and inclusive evaluation.

\section{Experimental Study}

In our work, we consider three different sets of experiments for training and evaluation: \textbf{1.} We establish an initial benchmark by training each dataset individually using all their original emotion classes. \textbf{2.} We use only four emotions (neutral, angry, happy, and sad) to train and test our SER models. We do this for both individual and combined settings. \textbf{3.} We use five emotions adding surprise to the previous four to train and test our SER models. We do this for both individual and combined settings.

We used a consistent training setup involving 5-fold cross-validation. This means we divided the data so that each fold used a different speaker for validation and the rest for training. For datasets with less than five speakers, we adjusted the number of folds to match the number of speakers. Additionally, with the TESS dataset, which has only two speakers, we used one for testing and the other for training, applying 5-fold cross-validation on the training speaker's data. In the case of the EmoFilm dataset, we used 85\% of the data for training, still following the 5-fold cross-validation method. For combined datasets, however, we adapted 5-fold cross-validation based on percentage splits across the dataset to ensure even distribution. It is important to note that we leave one speaker from each dataset as a test set right from the start, calculating a weighted average on the speaker-out test set across the 5-folds to ensure uniform and comprehensive evaluation metrics across all experiments.


\subsection{Individual and combined 4-emotion dataset evaluation}



For every dataset, we train a Whisper-based model using the four primary emotions: neutral, angry, happy, and sad. If a dataset lacks one of these emotions, we include from the dataset whichever available subset from these four emotions. \textbf{Table \ref{table:dataset_counts} } shows that we considered the most common 4 emotions across all datasets. For the combined dataset, we combined 11 SER datasets into a single comprehensive dataset based on the same 4 emotions. We included audio clips with duration ranging from 2 to 13 seconds, to ensure consistency and comparability across the combined dataset. Speech clips within this duration range are the most common duration of data across all datasets.
After combining emotions from various datasets, the data distribution exhibited imbalance, as illustrated in \textbf{Table \ref{fig:imbalance}}. This imbalance in the distribution of emotional categories across the combined dataset could potentially influence both the training and evaluation phases of our model. Explored four techniques to determine if any could improve model generalization: no sampling, randomly downsampling to the lowest category, and oversampling performed using the Synthetic Minority Over-sampling Technique (SMOTE) \cite{fernandez2018smote} and the Adaptive Synthetic Sampling Method for Imbalanced Data (ADASYN)\cite{he2008adasyn}. We applied the SMOTE and ADASYN algorithms to the audio data. The oversampling process was conducted to balance the distribution of emotion labels, ensuring that low-frequency emotions were over-sampled to match the frequency of the highest emotion category.

\begin{table}[!t]
\centering
\caption{Comparison of emotion distribution in combined 11 SER datasets with different sampling techniques.}
\label{tab:transposed_emotion_counts}
\resizebox{\columnwidth}{!}{%
\begin{tabular}{l|cccc|cccc}
\hline
 & \multicolumn{4}{c|}{\textbf{4-Emotions}} & \multicolumn{4}{c}{\textbf{5-Emotions}} \\
\textbf{Emotion/Method} & \textbf{Original} & \textbf{Undersamp} & \textbf{SMOTE} & \textbf{ADASYN} & \textbf{Original} & \textbf{Undersamp} & \textbf{SMOTE} & \textbf{ADASYN} \\ \hline
\textbf{Neutral (neu)}               & 10044    & 6157         & 10044 & 10044   & 10044    & 4056         & 10044 & 10044   \\
\textbf{Angry (ang)}                & 7427     & 6157         & 10044 & 10044   & 7427     & 4056         & 10044 & 10044   \\
\textbf{Happy (hap)}                  & 6286     & 6157         & 10044 & 10044   & 6286     & 4056         & 10044 & 10044   \\
\textbf{Sad (sad)}                    & 6157     & 6157         & 10044 & 10044   & 6157     & 4056         & 10044 & 10044   \\
\textbf{Surprised (sur)}              & -        & -            & -     & -       & 4056     & 4056         & 10044 & 10044   \\ \hline
\end{tabular}%
}
\label{fig:imbalance}
\end{table}

\subsection{Individual and combined 5-emotions dataset evaluation}
This step was taken to ascertain the model's robustness in handling an expanded set of emotion labels. we trained a Whisper-based model on a set of 5 emotions neutral, angry, happy, surprised, and sad. We train the model on each dataset using the same settings as in the individual training described in the previous experiment in Section 3.1. We repeat training and test settings in combined dataset training in the previous experiment, but with 5 emotions neutral, angry, happy, surprised, and sad.


\begin{figure*}[ht] 
  \centering
  \includegraphics[width=\textwidth]{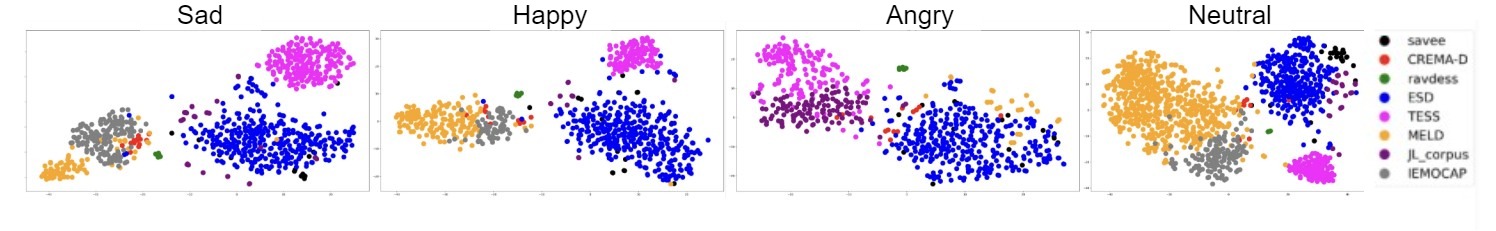} 
  \caption{t-SNE visualization of Whisper model's embeddings showcasing clusters of emotional speech from various datasets.}
  \label{fig:tsne}
\end{figure*}

\begin{table*}[ht]
\centering
\caption{Experiment Results for 4-Emotions, 5-Emotions, and N-Emotions, detailing performance across training criteria and data sampling techniques; Tr Co refers to training combination of datasets, Tr. sep. refers to training datasets separate, WA as weighted average, DS as downsampling, UN as original samples, SM refers to SMOTE sampling and AD as ADASYN oversampling}
\label{tab:experiment_results_4_5_n_emotions}
\renewcommand{\arraystretch}{1.3}
\setlength\tabcolsep{2pt} 
{\tiny 
\begin{tabular}{|l|c|cccc|c|c|cccc|c|c|c|c|}
\hline
\multirow{3}{*}{\parbox{1.5cm}{\raggedright \textbf{Dataset Name}} }& \multicolumn{6}{c|}{\textbf{4-Emotions WA Test} (\%)} & \multicolumn{6}{c|}{\textbf{5-Emotions WA Test} (\%)} & \multicolumn{2}{c|}{\textbf{N-Emotions WA Test} (\%)} \\
\cline{2-15} 
 & \multirow{2}{*}{\textbf{Emo. No.}} & \multicolumn{4}{c|}{\textbf{Tr. Co.}} & \multirow{2}{*}{\textbf{Tr. Sep.}} & \multirow{2}{*}{\textbf{Emo. No.}} & \multicolumn{4}{c|}{\textbf{Tr. Co.}} & \multirow{2}{*}{\textbf{Tr. Sep.}} & \multirow{2}{*}{\textbf{Emo. No.}} & \multirow{2}{*}{\textbf{Tr. Sep.}} \\
\cline{3-6} \cline{9-12}
 &  & \textbf{DS} & \textbf{UN} & \textbf{SM} & \textbf{AD} &  &  & \textbf{DS} & \textbf{UN} & \textbf{SM} & \textbf{AD} &  &  & \\
\hline
\textbf{ESD} & 4 & \textbf{79.11} & 77.5 & 72.73 & 74.45 & 76.85 & 5 & 78.81 & 74.45 & 74.72 & 74.72 & 72.04 & 5 & 72.04 \\
\textbf{MELD} & 4 & 51.20 & 54.72 & 56.11 & 56.65 & \textbf{63} & 5 & 47.66 & 54.72 & 47.68 & 47.68 & 56.87 & 7 & 50 \\
\textbf{IEMOCAP} & 4 & 67.75 & 70.00 & 68.79 & 68.96 & \textbf{72.43} & 5 & 66.37 & \textbf{70.00} & 67.42 & 67.42 & 69.5 & 7&  56.1 \\
\textbf{CREMA-D }& 4 & 93.50 & \textbf{98.00} & 95 & 93.5 & 90.41 & 4 & 91 & \textbf{98.00} & 95.50 & 95.50 & 90.41 & 6 & 80 \\
\textbf{EMOV\_DB} & 2 & 99.45 & 99.64 & \textbf{99.7} & 99.68 & 94.4 & 2 & 99.6 & 99.64 & 99.66 & 99.66 & 94.4 & 5 & 80.9 \\
\textbf{ASVP-ESD} & 3 &  79.11 & 77.75 & 81.25 & 78.75 & \textbf{84.65} & 4 & 73.37 & 77.75 & 76.30 & 76.30 & 82.35 & 7 & 73.33 \\
\textbf{TESS} & 4 & \textbf{77.35} & 71.08 & 74.9 & 74.07 & 66.95 & 5 & 62.5 & 71.08 & 55.83 & 55.83 & 56.62 & 7 & 55.52\\
\textbf{JL\_corpus} & 4 & 67.50 & 56.45 & \textbf{68.95} & 63.42 & 53.47 & 4 & 53.42 & 56.45 & 64.34 & 64.34 & 53.47 & 10 &  32.25\\
\textbf{RAVDESS} & 4 & 85.00 & 87.50 & 87.50 & 81.67 & 80.83 & 5 & 87.5 & 87.50 & \textbf{88.13} & \textbf{88.13} & 80.5 & 8 & 74.69 \\
\textbf{SAVEE} & 4 & 71.67 & 72.50 & \textbf{77.78} & \textbf{77.78} & 68.05 & 5 & 67.5 & 72.50 & 70.45 & 69.77 & 73.11 & 7 &  67.71 \\
\textbf{EmoFilm} & 3 &  67.50 & 64.17 & 62.5 & 58.53 & \textbf{82.2} & 3 & 68.33 & 64.17 & 57.5 & 60.83 & 82.2 & 5 & 53.37\\
\hline
\textbf{Mean} & - & 76 & \textbf{78.64} & 77.24 & 75.22 & 75.84 & - & 72.36 & \textbf{75.11} & 72.56 & 73.38 & 73.74 & - & 63.20  \\
\hline
\end{tabular}
}
\vspace{-1em}
\end{table*}

\section{Results and Discussion}

In this section, we provide a comprehensive overview of speech emotion recognition performance across different datasets, emotion categories, training criteria, and data sampling techniques, offering valuable insights for the generalization of speech emotion recognition. Table 3. illustrates the performance of the Whisper-based speech emotion recognition system developed under various combinations of dataset, and oversampling techniques. In general, the performance of SER varies across datasets and emotion categories. For instance, datasets like emov\_db consistently demonstrate high accuracy, while others like MELD exhibit lower performance under all conditions. Overall, using original data distribution or using SMOTE sampling tends to yield better performance compared to downsampling and ADASYN sampling. For instance, mean accuracy for training SER using a combination of 4 emotions demonstrates that using original data distribution and SMOTE sampling provides better results than other conditions. This suggests that maintaining the original sample distribution or using synthetic samples for minority classes can enhance model performance. Furthermore, training systems separately on each dataset for 4 emotion and 5 emotion experiments under performed in comparison to combining datasets. Thereafter, performance varies across datasets and emotion categories. For instance, datasets like emov\_db consistently showcase high accuracy, while others like MELD display comparatively lower performance. The choice of data sampling technique significantly influences emotion recognition performance. Techniques like SMOTE and ADASYN, which address the class imbalance, result in improved accuracy compared to simple downsampling. This highlights the importance of addressing class imbalance in training data to improve the robustness of emotion recognition models. 

Each dataset exhibits unique characteristics concerning the designing process of the dataset such as emotion label annotation, recording conditions of emotions, speaker demographics, and linguistic variations. etc, have an impact on emotion recognition performance. Hence, the factors such as data quality, diversity, and class distribution influence the effectiveness of trained models. For example, datasets like CREMA-D and emov\_db, which offer high-quality and diverse speech samples, tend to achieve better performance across different emotion categories. Therefore, it is vital to understand how different training conditions affect model generalization is essential for developing robust speech emotion recognition systems.
In conducting our research on 11 SER datasets with Whisper-based models, we encountered several limitations. 

Initially, the inherent diversity and imbalance in emotional representation across datasets introduced challenges in standardizing the training process, particularly when reducing the emotion categories to 4 and then to 5 specific emotions. Additionally, the process of combining datasets to improve model resilience encountered challenges in aligning emotional labels and intensities, which could potentially influence the consistency of our findings. The average accuracy findings for both the 4-emotion and 5-emotion experiments consistently demonstrated enhanced performance when utilizing the combined dataset, as opposed to separately training SER on each individual dataset. The comparison between models trained on individual datasets versus a combined dataset approach revealed variations in performance, suggesting that not all datasets contribute equally to model accuracy and generalizability.
t-SNE in \textbf{Figure \ref{fig:tsne}} visualizes the feature embeddings extracted from the Whisper model, representing various emotional states as discerned from multiple speech datasets. It's noteworthy how the datasets SAVEE, ESD, and JL\_corpus form closely knit clusters, suggesting that their emotional representations in the feature space are similar to each other. On the other hand, the MELD, IEMOCAP, and CREMA-D datasets tend to group closer to one another, indicating a different but consistent internal representation of emotions within these datasets. The RAVDESS and TESS datasets stand out with their unique positioning in the plot, reflecting distinct emotion representation patterns extracted by the Whisper model. Moreover, for each emotion exclusive representation on each dataset reflects the inherent variations in emotions across datasets and its impact on generalization of SER. 
\section{Conclusion}
This study proposes a novel approach to generalize SER model in leaving-one-speaker-out settings. Employing a leave-one-speaker-out methodology highlighted the model's robustness in real-world scenarios, facing speaker variability.  Extensive experiments were conducted by combining multiple SER datasets to train a Whisper-based model, followed by testing the model on a single speaker from each dataset, which yielded promising results, indicating a successful generalization across diverse datasets. 
These findings suggest that through careful dataset combination and targeted model training strategies, we can overcome some of the prevalent challenges in speech emotion recognition, paving the way for more universally applicable SER systems. In conclusion, our work yielded a generalized SER model adept at identifying emotions across a spectrum of datasets. This represents progress in the field, enabling the development of advanced emotion recognition systems capable of effectively handling diverse datasets. Future directions for emotion recognition in low-resource languages involve leveraging transfer learning and unsupervised learning approaches while considering cultural nuances and linguistic expertise.

\bibliographystyle{IEEEtran}
\bibliography{mybib}

\end{document}